\documentclass[prd,twocolumn,aps,showpacs,nofootinbib,nobibnotes,superscriptaddress,preprintnumbers]{revtex4}
\usepackage{epsfig}
\usepackage{graphics}
\usepackage{bm}
\usepackage{color}
\usepackage{dcolumn}
\usepackage{bm}
\usepackage{bbm}
\usepackage{amssymb}
\usepackage{amsmath}
\usepackage{latexsym}
\usepackage{float}
\usepackage{ifthen}
\usepackage{caption,subfig}
\usepackage{enumerate}
\usepackage{url}
\usepackage{caption,subfig}
\usepackage{amsopn}
\usepackage{hyperref}

\usepackage{amsfonts}
\usepackage{multirow}
\usepackage{array}
\usepackage{booktabs}
\usepackage{rotating}

\usepackage{ulem}
\def\clap#1{\hbox to 0pt{\hss#1\hss}}

\def\({\left(}
\def\){\right)}
\def\[{\left[}
\def\]{\right]}
\def\bea{\begin{eqnarray}}
\def\eea{\end{eqnarray}}
\def\be{\begin{equation}}
\def\ee{\end{equation}}
\def\ba{\begin{eqnarray}}
\def\ea{\end{eqnarray}}
\def\beq{\begin{eqnarray}}
\def\eeq{\end{eqnarray}}

\newcommand{\cs}{c_s}

\def\cs{c_{\rm s}}

\def\be{\begin{equation}}
\def\ee{\end{equation}}
\def\ba{\begin{eqnarray}}
\def\ea{\end{eqnarray}}
\def\beq{\begin{eqnarray}}
\def\eeq{\end{eqnarray}}

\def\L*{{\cal L}_*}
\def\L{\mathcal{L}}
\def\({\left(}
\def\){\right)}

\def\<{\langle}
\def\>{\rangle}

\def\cs2{c_{s}^{2}}

\def\be{\begin{equation}}
\def\ee{\end{equation}}
\def\ba{\begin{eqnarray}}
\def\ea{\end{eqnarray}}
\def\beq{\begin{eqnarray}}
\def\eeq{\end{eqnarray}}

\def\L*{{\cal L}_*}
\def\L{\mathcal{L}}
\def\({\left(}
\def\){\right)}

\def\<{\langle}
\def\>{\rangle}

\definecolor{hyperref}{RGB}{026,028,087}

\begin{document}

\title{Generalization of the 2-form interactions
}

\date{\today,~ $ $}


\author{Lavinia Heisenberg} \email{lavinia.heisenberg@phys.ethz.ch}
\affiliation{Institute for Theoretical Physics,
ETH Zurich, Wolfgang-Pauli-Strasse 27, 8093, Zurich, Switzerland}

\author{Georg Trenkler} \email{trgeorg@student.ethz.ch}
\affiliation{Institute for Theoretical Physics,
ETH Zurich, Wolfgang-Pauli-Strasse 27, 8093, Zurich, Switzerland}

\date{\today}

\begin{abstract}
We systematically construct derivative self-interactions for massless and massive 2-forms. There exists a no-go theorem in the literature for constructing Galileon-like Lagrangians in four dimensions for the 2-form with gauge invariance, the Kalb-Rammond field. The presence of non-minimal couplings strongly relies on the contraction with divergenceless tensors. In four dimensions these are the Einstein tensor and the double dual Riemann tensor. Even though they are divergenceless on their own, their combination ceases to be. In the case of massless 2-forms we are not able to establish non-minimal couplings of the 2-form to the gravity sector with second order equations of motion due to the impossibility of building consistent combinations of divergenceless tensors. Using the systematical construction in terms of the Levi-Civita tensor, we aim at constructing Galileon-like derivative self-interactions for the massive 2-form. Apart from $L_2$ and $L_4$ we are not able to construct further Galileon-like Lagrangians. For the massive case, an important non-minimal coupling between the 2-form and the double dual Riemann tensor arises, which receives additional support from the decoupling limit. Promoting the interactions in $L_4$ requires the presence of appropriate non-minimal couplings and we give concrete examples for this.
\end{abstract}


\maketitle

\section{Introduction}
Within infrared modified gravity theories, models based on higher dimensions have been extensively studied. One important representative is the DGP model \cite{Dvali:2000hr}, where our universe corresponds to a $D=4$ dimensional brane living in a 5 dimensional bulk. From the 4 dimensional point of view, the effective field theory contains a scalar field on top of the standard spin-2 field. One characteristic of this scalar field are its second order derivative interactions even though the field equations remain of second order. The latter is important for avoiding Ostrogradski instabilities. 

Soon, this unique property of the helicity-0 mode was generalized to the class of Galileon theories, which themselves became fashionable \cite{Nicolis:2008in}. In $D=4$ dimensions there is only a finite number of such interactions with explicitly second order equations of motion and invariance under a constant shift of the field and its gradient. The generalization of these interactions to include curvature effects led to the rediscovery of Horndeski theories \cite{Horndeski:1974wa}. They constitute the most general actions for a scalar-tensor theory with second order equations of motion.

One interesting follow-up question was, whether these Galileon-like Lagrangians can be constructed for arbitrary p-forms \cite{Deffayet:2010zh}. One immediate result stated the non-existence of massless vector Galileons in $D=4$ dimensions (see also \cite{Deffayet:2016von}). This no-go theorem does not extend to the case of massive spin 1 fields and one can construct non-trivial non-gauge invariant derivative self-interactions of the massive vector field with three propagating degrees of freedom, giving rise to generalized Proca theories \cite{Heisenberg:2014rta,Allys:2015sht,Jimenez:2016isa}. These theories have opened up interesting phenomenological applications in cosmology, astrophysics and black hole physics \cite{phenomeno} (see also \cite{Heisenberg:2018vsk}).

In \cite{Deffayet:2010zh} it was also shown that the construction of Galileon-like Lagrangians in $D=4$ dimensions for a massless 2-form, i.e. a Kalb-Rammond field fails. Starting from $D=7$ dimensions it is possible to write down Galileon interactions for a 2-form while maintaining explicitly the gauge invariance. In this Letter, we study the systematical construction of derivative self-interactions for both massless and massive 2-forms in 4 dimensions. This approach is highly based on the antisymmetric structure of the Levi-Civita tensors. 

It seems that the difficulty of constructing Galileon-like Lagrangians for the massless 2-form also persists for the massive 2-form interactions $L_3$, $L_5$ and $L_6$, at least using the aforementioned systematical construction. We will explicitly show the construction of interactions belonging to $L_2$ and $L_4$. Concerning the non-minimal couplings in curved spacetimes, we face difficulties to establish divergenceless tensors to which the gauge invariant strength tensor of the massless 2-form could couple. The massive case, however, admits a non-trivial coupling to the double dual Riemann tensor. We also show that promoting the $L_4$ interactions of the massive 2-form to curved spacetime requires the presence of adequate non-minimal couplings.

  
\section{Massless 2-forms}
In this section we consider first a massless antisymmetric Lorentz 2-form, which is often referred to as Kalb-Ramond field $B_{\mu\nu}$.
Lorentz invariance and masslessness imply that the theory describing the dynamics of the massless 2-form has to be invariant under the gauge transformation
\begin{equation} \label{eq1}
B_{\mu\nu} \rightarrow B_{\mu\nu} + \delta B_{\mu\nu}
\end{equation} 
with $\delta B_{\mu\nu} = \partial_{\mu}\epsilon_{\nu}-\partial_{\nu}\epsilon_{\mu}$ and $\epsilon_{\mu}$ being an arbitrary vector field.
Note that the gauge transformation has a redundancy $\epsilon_{\mu} \rightarrow \epsilon_{\mu}+\partial_{\mu}\lambda$,
where $\lambda$ is an arbitrary function.

In analogy to the massless spin-1 field, gauge invariance requires the 2-form to enter the Lagrangian in form of its corresponding field strength tensor
\begin{equation} \label{eq2}
H_{\mu\nu\rho} = \partial_{\mu} B_{\nu\rho} + \partial_{\nu} B_{\rho\mu} + \partial_{\rho} B_{\mu\nu} 
\,.
\end{equation}
The action for a massless antisymmetric 2-form can be written as
\begin{equation} \label{eq3}
S = -\frac{1}{12}\int d^{4}x H_{\mu\nu\rho}H^{\mu\nu\rho}\,.
\end{equation}
This action (\ref{eq3}) is clearly invariant under the gauge transformations given in equation (\ref{eq1}). It is precisely this gauge symmetry that ensures that the Kalb-Ramond field has only one propagating degree of freedom in 4 dimensions.

We can compute the equations of motion for the 2-form, yielding
\begin{align} \label{eq6}
\partial_{\rho}\Pi^{\rho\mu\nu} = 0 \qquad \Rightarrow  \qquad \partial^{\rho}\partial_{[\rho} B_{\mu\nu]} = 0\,,
\end{align}
where we denoted the antisymmetrization by square brackets and the conjugate momenta $\Pi^{\rho\mu\nu}$ of the 2-form by
\begin{align} \label{eq5}
\Pi^{\rho\mu\nu} 
= -\frac{1}{6}H^{\alpha\beta\gamma}\frac{1}{2}\delta_{\alpha}^{[\rho}\delta_{\beta}^{\mu}\delta_{\gamma}^{\nu]}
= -\frac{1}{2}H^{\rho\mu\nu} \,.
\end{align}
The equations of motion \eqref{eq6} rewritten in momentum space correspond to
\begin{equation} \label{eq8}
p^{2}B_{\mu\nu} + p^{\rho}(p_{\nu}B_{\rho\mu} - p_{\mu}B_{\rho\nu}) = 0 \,.
\end{equation}
We can introduce spacetime light-cone coordinates - a method widely used in string theory - to completely fix the gauge and count the independent propagating degrees of freedom of the Kalb-Ramond field. In these new coordinates Greek indices run over $\{+, -, i\}$, whereas Latin indices run over $\{1,...,D-1\}$. The $\epsilon_{+}$ component can be fixed to zero by choosing the arbitrary function $\lambda = - \frac{\epsilon_{+}}{p_{+}}$. The transformation of the 2-form on the other hand reads in momentum space
\begin{align} \label{eq10}
&\delta B_{+-} = p_{+}\epsilon_{-}, \qquad\qquad \;\;\;\; \delta B_{+i} = p_{+}\epsilon_{i}\,,\nonumber \\
&\delta B_{-i} = p_{-}\epsilon_{i} - p_{i}\epsilon_{-}, \quad \;\;\;\;\;\;\delta B_{ij} = p_{i}\epsilon_{j} - p_{j}\epsilon_{i} \,.
 \nonumber
\end{align}
It can be seen from the above equations that the components $B_{+-}$ and $B_{+i}$ can be removed for a suitable choice of $\epsilon_{\mu}$. 
With this gauge choice, the $\mu=+$ component of equation (\ref{eq8}) gives
\begin{equation} \label{eq11}
p^{2}\underbrace{B_{+\nu}}_{=0}+ p^{\rho}(p_{\nu}\underbrace{B_{\rho+}}_{=0} - p_{+}B_{\rho\nu})=0 \Rightarrow p_{+}p^{\rho}B_{\rho\nu} = 0\,,
\end{equation}
which leads to the generalization of the Lorentz gauge
\begin{equation} \label{eq12}
p^{\rho}B_{\rho\nu} = 0 \Rightarrow p^{l}B_{l\nu} = 0 
\end{equation}
in which the equations of motion for the 2-form field reduce to a wave equation.\\
Thus, we have 
\begin{align} \label{eq13}
& \frac{D(D-1)}{2}-\underbrace{1}_{B_{+-}=0}-\underbrace{(D-2)}_{B_{+i}=0}-\underbrace{(D-2)}_{p^{l}B_{l\nu}=0} \nonumber\\ 
& = \frac{1}{2}(D^{2}-5D)+3 \text{ degrees of freedom}
\end{align}
which for $D=4$ indeed becomes 1 degree of freedom. A massless 2-form in 4 dimensions propagates one physical degree of freedom.

We can construct a dual vector field $\bar{H}_{\mu}$ by contracting the field strength with the totally antisymmetric Levi-Civita symbol $\epsilon_{\mu\nu\rho\sigma}$ (with $\epsilon_{0123}=+1$) in 4 dimensions
\begin{equation} \label{eq14}
\bar{H}_{\mu} = \epsilon_{\mu\nu\rho\sigma}H^{\nu\rho\sigma}\,,
\end{equation}
which satisfies the Bianchi-like idendity $\partial^{\mu}\bar{H}_{\mu} = 0$ due to the total antisymmetry of the Levi-Civita symbol. Consequently, the equation (\ref{eq6}) can be written as
\begin{equation} \label{eq15}
\partial_{\rho}\bar{H}_{\mu}-\partial_{\mu}\bar{H}_{\rho} = 0
\end{equation} 
implying that $\bar{H}_{\mu}$ has to be the gradient of a scalar field $\Phi$, i.e $\bar{H}_{\mu} = \partial_{\mu}\Phi$.
Using this, the action in equation (\ref{eq3}) can be dually rewritten as
\begin{equation} \label{eq16}
S_{\Phi} = -\frac{1}{2}\int d^{4}x \partial_{\mu}\Phi\partial^{\mu}\Phi\,.
\end{equation}
From equation (\ref{eq16}) it is clear that the dynamics of a massless, antisymmetric 2-form $B_{\mu\nu}$ is equivalent to that of a scalar field $\Phi$.
\subsection{First order form}
The analysis of the propagating degree of freedom of the Kalb-Ramond field simplifies significantly in the first order form. By this, we mean treating $B_{\mu\nu}$ and $H_{\mu\nu\alpha}$ (or equivalently the conjugate momenta $\Pi^{\rho\mu\nu}$) as a priori independent fields, each satisfying a 1st order equation of motion. This framework is particularly useful when working in the Hamiltonian formulation of the classical theory, where the connection between the Hamiltonian density, the Lagrangian density and the conjugate momenta is given by a Legendre transformation.\\
Considering the action 
\begin{equation} \label{eq17}
S= -\frac{1}{12}\int d^{4}x \left(2H^{\mu\nu\alpha}\partial_{[\mu}B_{\nu\alpha]} - H^{\mu\nu\alpha}H_{\mu\nu\alpha}\right)
\end{equation}  
we can perform two independent variations with respect to $B_{\mu\nu}$ and $H_{\mu\nu\alpha}$ leading to the 1st order equations
\begin{equation} \label{eq18}
\frac{\delta S}{\delta B_{\nu\alpha}} \rightarrow \partial_{\rho}H^{\rho\mu\nu} = 0
\end{equation}
and 
\begin{equation} \label{eq19}
\frac{\delta S}{\delta H^{\mu\nu\alpha}} \rightarrow H^{\mu\nu\alpha} = \partial_{[\mu}B_{\nu\alpha]} \,.
\end{equation}
We see that taken together, equations (\ref{eq18}) and (\ref{eq19}) are consistent with the results derived in the Lagrangian formulation. Plugging them into equation (\ref{eq17}) we would obtain the standard action in \eqref{eq3}.
Moreover, equation (\ref{eq18}) implies 
\begin{equation} \label{eq20}
H^{\mu\nu\alpha} = \epsilon^{\mu\nu\alpha\beta}\partial_{\beta}\Phi
\end{equation}
where $\Phi$ is again a scalar field. Plugging this into the action in equation (\ref{eq3}), we obtain the action of a massless scalar field given in equation (\ref{eq16}) and hence confirm also in the first order formalism that there is indeed only 1 propagating degree of freedom.
\subsection{Galileon interactions for the massless 2-form}
We have seen that the massless antisymmetric 2-form propagates only one physical dof. A natural question arises as to whether other non-trivial derivative self-interactions could be constructed for the massless 2-form, without altering the number of propagating degrees of freedom. This question led immediately to a no-go theorem in \cite{Deffayet:2010zh}, where it was shown that one cannot construct Galilean interactions for a massless 2-form field in $D=4$ dimensions. In fact, allowing higher dimensions (starting from $D=7$) one can construct Galileon interactions of the type
\begin{equation}
L^{(D=7)}=\epsilon^{\mu\nu\rho\sigma\tau\phi\chi}\epsilon^{\alpha\beta\gamma\delta\epsilon\xi\eta}H_{\mu\nu\rho}H_{\alpha\beta\gamma}\partial_\sigma H_{\delta\epsilon\xi}\partial_\eta H_{\tau\phi\chi} \,.
\end{equation}
However, in 4 dimensions one cannot construct Galileon interactions with gauge symmetry for the 2-form, leaving the kinetic term as the unique gauge invariant term. A similar no-go theorem exists for the massless spin-1 field, where one cannot extend interactions of the form $(\partial F)^n$ beyond the Maxwell kinetic term \cite{Deffayet:2013tca}, while keeping the gauge invariance.

\subsection{Non-minimal couplings for the massless 2-form}
We have mentioned in the previous subsection that there exist a no-go theorem for derivative self-interactions for the massless 2-form as it is the case for a massless spin-1 field. Nevertheless, in the latter case one can construct a non-minimal coupling of the massless spin-1 field to the double dual Riemann tensor $L^{\alpha\beta\gamma\delta}F_{\alpha\beta}F_{\gamma\delta}$. We would like to understand whether similar non-minimal couplings exist in the case of the massless 2-form. We will demand that
\begin{itemize}
\item all terms should maintain the gauge symmetry, meaning that the 2-form $B_{\mu\nu}$ has to enter the action only through its manifestly gauge-invariant field strength $H_{\mu\nu\rho}$,
\item the equations of motion for the involved fields $g_{\mu\nu}$ and $B_{\mu\nu}$ are at most of second order\,,
\item the massless 2-form only propagates one degree of freedom while we keep the two degrees of freedom in the gravity sector.
\end{itemize}
In order to guarantee the last two conditions,  the field strength $H_{\mu\nu\rho}$ can only couple to the divergence-free tensors: the metric $g^{\alpha\beta}$, the Einstein tensor $G^{\alpha\beta} = R^{\alpha\beta} - \frac{1}{2}R g_{\mu\nu}$ and the double dual Riemann tensor $L^{\alpha\beta\gamma\delta} = \frac{1}{4}\epsilon^{\alpha\beta\mu\nu}\epsilon^{\gamma\delta\rho\sigma}R_{\mu\nu\rho\sigma}$. They are crucial in order not to have higher order equations of motion as a consequence of partial integration and hence acting additional covariant derivatives on them coming from the field strength part. Note, that these tensors are only divergence-free on their own, but not combinations thereof such as $(gG)$, $(GG)$, $(gL)$, $(GL)$, etc.
Indeed, all possible contractions give rise to such tensor combinations and the resulting index structure prevents us from actually using their being divergenceless as we will see below. 

Needless to say, that several contractions vanish identically on grounds of symmetry properties of the three divergenceless tensors and the totally antisymmetric nature of the field strength tensor $H_{\mu\nu\rho}$, for instance $H_{\mu\nu\rho}H_{\alpha\beta\gamma}g^{\nu\rho}L^{\alpha\beta\gamma\mu} = 0$ and $H_{\mu\nu\rho}H_{\alpha\beta\gamma}G^{\nu\rho}L^{\alpha\beta\gamma\mu} = 0$, etc.
Furthermore, couplings through only $g^{\mu\nu}$ are for obvious reasons allowed: tensors built from $g^{\mu\nu}$ without any derivatives can be contracted with $H$'s without any care, such as the standard kinetic term or any non-linear function thereof $f(H)$. However, since the Einstein tensor and the double dual Riemann tensor carry two derivatives acting on the metric, non-minimal couplings containing them are more problematic. Unfortunately, we were not able to construct any contraction that would maintain their divergeceless nature, as we discuss below through concrete examples.\\

\textbf{Couplings through $G^{\mu\nu}$:} We immediately see that due to $G$ being symmetric and carrying 2 indices, we need at least 2 $H$'s. Consider a term of type $(ggG)$:
\begin{align} \label{eq27}
&H_{\mu\nu\alpha}H^{\mu\nu\beta}G^{\alpha}_{\;\beta} \supset \nabla_{\mu}B_{\nu\alpha}\nabla^{\mu}B^{\nu\beta}g^{\alpha\lambda}G_{\lambda\beta}\nonumber \\
&\sim B_{\nu\alpha} [(\nabla_{\mu}\nabla^{\mu}B^{\nu\beta})g^{\alpha\lambda}G_{\lambda\beta}
+\nabla^{\mu}B^{\nu\beta}(\nabla_{\mu}g^{\alpha\lambda})G_{\lambda\beta} \nonumber\\
&+\nabla^{\mu}B^{\nu\beta}g^{\alpha\lambda}(\nabla_{\mu}G_{\lambda\beta})] \,.
\end{align}
As one can clearly see from above example potentially problematic terms $\partial^3g$ arise that would yield higher order equations of motion. Even though $G$ is divergenceless on its own, the combination $(ggG)$ with two inverse metric and one Einstein tensors is not anymore. The same happens for terms including more $G$'s like for instance $H_{\mu\alpha\rho}H_{\nu\beta\sigma}G^{\mu\nu}G^{\alpha\beta}g^{\rho\sigma}$ and $H_{\mu\alpha\rho}H_{\nu\beta\sigma}G^{\mu\nu}G^{\alpha\beta}G^{\rho\sigma}$ since they bring along $\nabla_{\mu}G^{\rho\sigma}\neq 0$. There, the multiplications $(GGg)$ and $(GGG)$ acquire a non-vanishing divergence, and we lose the necessary fundamental property in order to satisfy the last two conditions mentioned above.\\

\textbf{Couplings through $L^{\mu\nu\alpha\beta}$:} Since the double dual Riemann tensor has four indices, we need at least two $H$'s to construct fully contracted objects at the price of introducing an inverse metric or the Einstein tensor. This on the other hand faces the difficulty that $(gL)$ and $(GL)$ are not divergenceless anymore. For instance,
\begin{align} \label{eq30}
L^{\mu\nu\alpha\beta}H_{\mu\nu\rho}H_{\alpha\beta\sigma}g^{\rho\sigma} 
&\supset L^{\mu\nu\alpha\beta}(\nabla_{\rho}B_{\mu\nu})H_{\alpha\beta\sigma}g^{\rho\sigma} \\
&\sim\underbrace{(\nabla_{\rho} L^{\mu\nu\alpha\beta})}_{\neq 0\rightarrow \text{problematic}}B_{\mu\nu}H_{\alpha\beta\sigma}g^{\rho\sigma}\nonumber
\end{align}
and similarly
\begin{align} \label{eq31}
L^{\mu\nu\alpha\beta}H_{\mu\nu\rho}H_{\alpha\beta\sigma}G^{\rho\sigma} 
\supset\underbrace{(\nabla_{\rho} L^{\mu\nu\alpha\beta})}_{\neq 0\rightarrow \text{problematic}}B_{\mu\nu}H_{\alpha\beta\sigma}G^{\rho\sigma}\,.
\end{align}
Hence, couplings through the double dual Riemann tensor face the same challenge as those through the Einstein tensor. 

One could wonder if interactions involving the dual-field $\bar{H}$, e.g of type $(G\bar{H}\bar{H})$ or $(LH\bar{H})$ could represent valid interactions. However, these types of interactions are either parity-violating (which we do not consider here) or they can be identified as couplings of the dual field to the Riemann tensor which in turn can be related to couplings of the double dual Riemann tensor to the field strength. In this sense, they do not represent any new interaction that has not already been considered in the above discussion.

Based on the argument that the absence of higher order equations of motion and the propagation of $1+2$ degrees of freedom in the presence of non-minimal couplings strongly rely on the realization of divergenceless tensors, we have seen in this subsection, that it is impossible to contract the gauge invariant field strength of the 2-form with a divergenceless tensor. This comes hand in hand with the difficulty, that even though the inverse metric, the Einstein tensor and the double dual Riemann tensor are divergence-free on their own, the products thereof are not. 

This might seem to contradict the results presented in \cite{Yoshida:2019dxu}, where a non-minimal coupling for a gauge invariant 2-form was constructed using a duality relation to a subclass of Horndeski theories. The advocated interaction is of the form
\begin{equation}\label{nonMinC3G2H}
\frac{1}{\det G}G^{\rho\alpha}G^{\sigma\beta}G^{\lambda\gamma}H_{\rho\sigma\lambda}H_{\alpha\beta\gamma}\,.
\end{equation}
Based on above argument, we had classified terms of the form $GGGHH$ as being problematic since the combination of 3 Einstein tensors acquires a non-vanishing divergence. Bear in mind that this interaction \eqref{nonMinC3G2H} is just a complicated way of rewriting the original Horndeski interaction $G^{\mu\nu}\partial_\mu\Phi \partial_\nu\Phi$ by means of a non-trivial duality relation that carries an inverse Einstein tensor. One way of reconciling this finding in \cite{Yoshida:2019dxu} with our results is that in terms of an expansion of this interaction in terms of curvature the truncation at a given order will encounter the problems associated to the non-vanishing divergence mentioned above but the summation of this infinite series as an inverse Einstein tensor (or inverse determinant thereof) might then circumvent the problem \footnote{We thank Jose Beltran Jimenez and Daisuke Yoshida for discussions on this. See also \cite{Jimenez:2019hpl} and \cite{Gabadadze:2012tr} for a related discussion.}.

\section{Massive 2-forms}
	In $D=4$ dimensions there is a no-go result for derivative self-interactions for a massless 2-form \cite{Deffayet:2010zh}. One might wonder, whether this no-go result can be avoided by abandoning the associated gauge invariance of the 2-form, in the same spirit as the generalized Proca theories. In this section, we will consider massive antisymmetric 2-forms and study their allowed interactions using the systematical construction scheme via the Levi-Civita tensor. 
In the previous section we have seen that a massless 2-form propagates only one physical degree of freedom. If we explicitly break the gauge invariance, the massive case contains three degrees of freedom.

\subsection{First order form}
The propagation of three modes becomes quickly transparent in the first order formalism.
Consider the following action
\begin{align}\label{Lmassive2form}
S= -\frac{1}{12}\int d^{4}x \Big(&2H^{\mu\nu\alpha}\partial_{[\mu}B_{\nu\alpha]} - H^{\mu\nu\alpha}H_{\mu\nu\alpha} \nonumber\\
&+2m^{2}B_{\mu\nu}B^{\mu\nu}\Big)\,,
\end{align}
where $H_{\mu\nu\rho}$ and $B_{\mu\nu}$ are again treated as two independent fields. Note the presence of a mass term, which explicitly breaks the previous gauge symmetry of the massless 2-form.
Variation with respect to the field strength $H_{\mu\nu\rho}$ gives again
$H_{\mu\nu\rho}=\partial_{[\mu}B_{\nu\rho]}$.
Contrary to the massless case, the Lagrangian \eqref{Lmassive2form} is no longer linear in the B-field preventing it from acting as a Lagrange multiplier.
Variation with respect to the 2-form $B_{\mu\nu}$ results in
\begin{equation}
\partial_{\alpha}H^{\alpha\mu\nu}-m^{2}B^{\mu\nu}=0\,,
\end{equation}
which we can solve for $B ^{\mu\nu}$. Plugging the result back into the action yields
\begin{equation}
S= -\frac{1}{12}\int d^{4}x \left(-\frac{4}{m^{2}}\partial_{\mu}H^{\mu\nu\alpha}\partial^{\lambda}H_{\lambda\nu\alpha} - H^{\mu\nu\rho}H_{\mu\nu\rho}\right)\,.
\end{equation}
We can dualize the action in terms of the Hodge dual for $A$ by
\begin{equation}
H^{\mu\nu\alpha}=\epsilon^{\mu\nu\alpha\beta}A_{\beta}\,,
\end{equation}
which gives the dual action of a massive vector field after insertion
  \begin{equation}
S[A] = \int d^{4}x \left(-\frac{1}{4}F_{\mu\nu}(A)F^{\mu\nu}(A)-\frac{1}{2}A_{\mu}A^{\mu} \right) \,.
\end{equation}
This duality relation shows that the massive antisymmetric 2-form propagates the same number of physical degrees of freedom as a massive Proca field, namely 3. This effect is sometimes called in string theory spin jumping of the Kalb-Ramond field from spin 0 to spin 1.

\subsection{Galileon interactions for massive 2-forms}\label{GalMass2form}
In this subsection, we are after Galileon-type interactions for the massive 2-form. For this purpose, we will follow a procedure that has been extensively exploited in the literature, that is based on using the antisymmetry properties of the Levi-Civita tensor. We will demand that the equations of motion remain second order despite the presence of higher order derivative interactions in the Lagrangian and that the massive 2-form propagates only 3 degrees of freedom.

The first immediate interactions can be constructed by potential-like and gauge-invariant terms in form of
\begin{equation} \label{LagL2}
L_{2} = f_{2}(B_{\mu\nu}, H_{\mu\nu\rho}, \bar{H}_{\mu})\,.
\end{equation}
Since they contain either gauge-invariant contractions or terms without derivatives acting on the 2-form, they trivially satisfy our two conditions.
As next, we can perform a series expansion order by order in powers of the fundamental object $\partial_{\alpha}B_{\mu\nu}$ contracted with the Levi-Civita tensors, in the schematic form $ f(B^2)\epsilon\epsilon(\partial B)^{m}B^{n}$, where $m, n \in {0, 1, 2, ...}$ etc. 

The attempt to construct interactions with $m=1$, i.e. linear in $\partial B$, fails immediately since the Levi-Civita tensor carries four indices whereas our fundamental object $\partial_\alpha B_{\mu\nu}$ carries three indices. This does not alter by including $n\ne0$, since the odd number of remaining uncontracted indices cannot be compensated by the even number of indices in $B_{\mu\nu}$. Therefore, we have for the cubic Lagrangian
\begin{equation} \label{LagL2}
L_{3} = 0\,.
\end{equation}
For the next order, for $m=2$, we have an even number of indices contracted with the Levi-Civita tensors, so that one can start constructing non-trivial terms. We can for instance consider the following three possible contractions
\begin{align} 
L_{4}^{(0B)}=L_{4}^H+L_{4}^R+L_{4}^T 
\end{align} 
where 
\begin{align} 
L_{4}^H &= \epsilon^{\mu\nu\rho\sigma}\epsilon^{\alpha\beta\gamma}_{\;\;\;\;\;\;\sigma}\partial_{\mu}B_{\nu\rho}\partial_{\alpha}B_{\beta\gamma}  \\
L_{4}^R &=\epsilon^{\mu\nu\rho\sigma}\epsilon^{\alpha\beta\gamma}_{\;\;\;\;\;\;\sigma}\partial_{\mu}B_{\nu\beta}\partial_{\alpha}B_{\rho\gamma}  \\
L_{4}^T &=\epsilon^{\mu\nu\rho\sigma}\epsilon^{\alpha\beta\gamma}_{\;\;\;\;\;\;\sigma}\partial_{\mu}B_{\alpha\rho}\partial_{\nu}B_{\beta\gamma} \,.
\end{align} 
The first contraction $L_{4}^H$ gives rise to a contribution proportional to the kinetic term 
\begin{equation}
L_{4}^H=-\frac23H_{\mu\nu\rho}H^{\mu\nu\rho}
\end{equation}
and hence can be simply absorbed into $L_{2}$. The second contribution is not independent and can be entirely expressed in terms of the first and third contractions
\begin{equation}
L_{4}^R=\frac12(L_{4}^H-L_{4}^T)\,.
\end{equation}
Hence, the genuinely new interaction comes only from $L_{4}^T$. Therefore, we will disregard the other two for now. It corresponds to
\begin{align}
L_{4}^T&=\epsilon^{\mu\nu\rho\sigma}\epsilon^{\alpha\beta\gamma}_{\;\;\;\;\;\;\sigma}\partial_{\mu}B_{\alpha\rho}\partial_{\nu}B_{\beta\gamma}\nonumber\\
&=\partial_\mu B^{\mu\nu}\partial_\alpha B_\nu{}^{\alpha}+\partial_\nu B_{\mu\alpha}\partial^\alpha B^{\mu\nu}\,.
\end{align}
This term is quite special. On its own standing like this, it does not contribute to the equations of motion $\frac{\delta L_{4}^T}{\delta B_{\mu\nu}}=0$, so it represents a total derivative. In fact, on closer inspection one actually recognizes that the second term is identical to the first term but with a minus sign, after integrating by parts twice.

However, it does not correspond to a total derivative once it is multiplied with an overall function $f_4(B^2)$.
Hence, the construction at order $L_{4}$ gives rise to the genuinely new term, 
\begin{equation}\label{modKinL4}
L_{4}^{0B}=f_4(B^2)\Big(\partial_\mu B^{\mu\nu}\partial_\alpha B_\nu{}^{\alpha}+\partial_\nu B_{\mu\alpha}\partial^\alpha B^{\mu\nu} \Big)\,,
\end{equation}
that would go beyond $L_{2}$. This term looks like a modified kinetic term without gauge invariance multiplied by an overall function of the 2-form norm.

By replacing the inverse Minkowski metric that contracts the last indices in the Levi-Civita tensors by a 2-form field
one can also construct the $n=1$ term. The first two contractions in $L_{4}^R$ and $L_{4}^H$ vanish exactly in this case due to antisymmetric nature of $B_{\mu\nu}$. The only surviving term for $m=2$ and $n=1$ is
\begin{align} 
L_{4}^{(1B)}=
 \epsilon^{\mu\nu\rho\sigma}\epsilon^{\alpha\beta\gamma\delta}\partial_{\mu}B_{\alpha\rho}\partial_{\nu}B_{\beta\gamma} B_{\sigma\delta}\,,
\end{align} 
which can also be multiplied by a function of the 2-form norm. 

At this stage it can be noted that all Lorentz-indices belonging to the Levi-Civita symbols are already fully contracted meaning that all terms involving higher powers of $B_{\mu\nu}$ necessarily have to be contracted among themselves, such as
\begin{align} 
L_{4}^{(2B)}=
 \epsilon^{\mu\nu\rho\sigma}\epsilon^{\alpha\beta\gamma\delta}\partial_{\mu}B_{\alpha\rho}\partial_{\nu}B_{\beta\gamma} B_{\sigma\lambda}B^\lambda{}_\delta\,.
\end{align}
Note, that since our systematical construction $f(B^2)\epsilon\epsilon(\partial B)^{m}B^{n}$ is such that the derivatives acting on the 2-form are contracted with the two Levi-Civita tensors, we cannot grasp terms where the indices of $(\partial B)^{m}$ are contracted among themselves.

For contributions with $m=3$, we immediately observe that the eight indices of the two Levi-Civita tensors are not sufficient to contract the indices with $(\partial B)^3$, which contains nine indices. Hence, 
\begin{equation}
L_{i}=0 \qquad \text{for} \qquad i\geqq5
\end{equation}
Therefore, the systematical construction stops here and we are not able to construct Galileon interactions for the massive 2-form that would go beyond $L_{4}$. The term we constructed for $L_{4}$ in \eqref{modKinL4} might be called a {\it modified kinetic term}, that contributes in a non-trivial way only in the presence of an overall function of the 2-form norm. 

The interactions arising in $L_4$ can be better understood from the decoupling limit using the Stueckelberg trick. The broken gauge symmetry of the 2-form can be restored by introducing a massless spin-1 field as the Stueckelberg field $B_{\mu\nu} \to B_{\mu\nu}+\partial_{[\mu}A_{\nu]}$. In this way the massive 2-form gets decomposed into a massless 2-form and a massless spin-1 field (still 3 propagating degrees of freedom in total). In the decoupling limit where the mass of the 2-form is sent to zero but leading contributions are kept intact, we know that the pure massless 2-form and the massless spin-1 sectors cannot contain any Galileon-like interactions (due to existing no-go theorems). For instance, there cannot be any interaction of the form $f_4(F^2)\partial F\partial F$, where $F$ is the field strength tensor of the massless spin-1 field. The only way the interactions can contribute is via a mixing between the massless 2-form and the massless spin-1 field. And this is exactly the way how the above interactions in $L_4$ contribute. 

\subsection{Non-minimal couplings for the massive 2-form and curved spacetime}
In the case of massless a 2-form we were not able to construct non-minimal couplings due to the difficulty of finding a divergenceless tensor with the right index structure. Gauge invariance required that the 2-form only appears via the field strength tensor $H_{\mu\nu\alpha}$ and its contraction demanded the multiplication of divergenceless tensors, which is not divergenceless anymore. 

In the massive case, we cannot consider derivative non-minimal couplings either since we were not able to construct Galileon interactions beyond $(\partial B)^2$ in the previous subsection, at least using our systematical construction in four dimensions. It means that we cannot couple $\partial B$ to $G$ and $L$ since the corresponding Galileon derivative interaction is missing in order to compensate it. However, due to the broken gauge symmetry, we can consider contractions of the divergenceless tensors directly with the massive 2-form. Due to the antisymmetric nature of the 2-form, we cannot couple it to the Einstein tensor, $G^{\mu\nu}B_{\mu\nu}=0$. Based on the allowed symmetries, naively we could construct terms of the form $G^{\mu\nu}G^{\alpha\beta}B_{\mu\alpha}B_{\nu\beta}$. Even though the Einstein tensor is divergenceless, the multiplication of two Einstein tensors is not anymore. Hence, this term would give rise to higher order equations of motion for the metric, if we applied the derivatives of one $G$ to the other $G$ by integration by parts. Since it contains four derivatives, for its compensation we would need the presence of Galileon interactions at $(\partial B)^4$ order, which we were not able to construct using the systematical construction in terms of the Levi-Civita tensors in 4 dimensions.

Thus, the only possible non-minimal unique coupling is via the double dual Riemann tensor
\begin{equation}\label{mass2formLBB}
L=\sqrt{-g}L^{\mu\nu\alpha\beta}B_{\mu\nu}B_{\alpha\beta}\,.
\end{equation}
It would be interesting to study the implications of this non-minimal coupling in more detail. This unique non-minimal coupling is harmless on its own. However, if one were tempted to multiply it with an overall function of the 2-form norm, then this would need to be compensated by an appropriate interaction of the form $B^2(\partial B)^2$, as in the case of Horndeski and generalized Proca theories. 

This unique non-minimal coupling \eqref{mass2formLBB} finds additional support from the decoupling limit using the Stueckelberg field. If we rewrite the theory in terms of a massless 2-form and a massless spin-1 field via $B_{\mu\nu} \to B_{\mu\nu}+\partial_{[\mu}A_{\nu]}$, then we know that the only way how the massless spin-1 field can couple non-minimally to the gravity sector is through the unique interaction $L^{\alpha\beta\gamma\delta}F_{\alpha\beta}F_{\gamma\delta}$. Therefore, it is not possible to find any other non-minimal coupling beyond \eqref{mass2formLBB} that is fine on its own without spoiling the allowed coupling in the massless spin-1 field sector.

The interactions, that we constructed in \ref{GalMass2form} using the systematical construction need a special care when promoting them to curved spacetime. This is a difficult task, since it is not enough to guarantee the second order nature of the equations of motion. Even if the curved spacetime version would not yield higher order equations of motion, the gauge modes in the gravity sector and the non-physical modes in the massive 2-form should not propagate, i.e. we should still have 2+3 degrees of freedom. We give a few examples below but a rigorous analysis is out of scope of this Letter.

For the interactions in $L_{2}$ there is no harm in simply replacing the partial derivatives by covariant derivatives. Hence, we can straightforwardly write them as
\begin{equation} \label{LagL2curved}
L_{2} = \sqrt{-g}f_{2}(B_{\mu\nu}, H_{\mu\nu\rho}, \bar{H}_{\mu})\,.
\end{equation}
However, the non-trivial interaction $L_{4}^T$ in curved spacetime requires the presence of a non-minimal coupling to the Ricci scalar in order not to render the gauge modes of the massless spin-2 field dynamical. The analysis for cosmological backgrounds yields the following relative tuning as a possible compensation
\begin{align}\label{L4Tcurved}
L_{4}=&\sqrt{-g}\Big\{f_4(X)R \nonumber\\
&+3f_{4,X}\Big(\nabla_\mu B^{\mu\nu}\nabla_\alpha B_\nu{}^{\alpha}+\nabla_\nu B_{\mu\alpha}\nabla^\alpha B^{\mu\nu} \Big) \Big\}\,,
\end{align}
where $X=B^2$. As we mentioned above, the special non-minimal coupling \eqref{mass2formLBB}
$\sqrt{-g}L^{\mu\nu\alpha\beta}B_{\mu\nu}B_{\alpha\beta}$ is fine own its own. However, including an overall function of the 2-form norm in front would need the simultaneous presence of appropriate interactions of the form $B^2(\partial B)^2$. Due to the symmetries of our chosen background we could not find a decisive answer for such interactions. 
A Hamiltonian analysis of an arbitrary background with less symmetries in the ADM decomposition would yield more insight into the number of propagating degrees of freedom and the right relative tunings.

\section{Conclusion}
In this work we have studied the systematical construction of derivative self-interactions for massless and massive 2-forms together with their non-minimal couplings to gravity. Using the construction based on divergenceless tensors we were not able to build non-minimal couplings for the massless 2-form. We encountered also difficulties to construct Galileon-like interactions beyond $L_2$ and $L_4$ for the massive 2-form using a systematical construction based on the antisymmetric Levi-Civita tensor. The resulting possible interactions are not higher in derivatives than $(\partial B)^2$. The failure of constructing higher order derivative Galileon-like interactions comes hand in hand with the absence of derivative non-minimal couplings. A coupling between the 2-form and the double dual Riemann tensor emerges as the unique possible coupling that guarantees the second order nature of equations of motion. Unless the interactions in $L_2$, those constructed in $L_4$ need a special care while promoting them to curved spacetime. We gave some concrete examples for maximally symmetric backgrounds. It would be very interesting to study the cosmological and astrophysical implications of above constructed interactions for the massive 2-form, which we leave for a future work.

\acknowledgments 
We would like to thank Matthias Bartelmann, Jose Beltran Jimenez, Shinji Tsujikawa and Daisuke Yoshida for useful discussions.
LH is supported by funding from the European Research Council (ERC) under the European Unions Horizon 2020 research and innovation programme grant agreement No 801781 and by the Swiss National Science Foundation grant 179740.

\newpage

\end{document}